\documentclass[%
superscriptaddress,
 amsmath,amssymb,
 aps,
twocolumn
]{revtex4-1}

\usepackage{graphicx}
\usepackage{amssymb}
\usepackage{float}
\usepackage{amsmath}
\usepackage{verbatim}
\usepackage{graphicx}
\usepackage[toc,page]{appendix}
\usepackage{braket}
\usepackage{amsmath}
\usepackage{siunitx,esvect}

\DeclareMathOperator{\sinc}{sinc}
\usepackage{siunitx}
\usepackage{textcomp}
\usepackage{gensymb}
\usepackage{appendix}
\usepackage{braket}
\usepackage{upgreek}
\usepackage[breaklinks=true,colorlinks=true,linkcolor=blue,urlcolor=blue,citecolor=blue]{hyperref}
\usepackage{dcolumn}
\usepackage{bm}
\usepackage{natbib}
\usepackage{multirow}
\newcommand{\parallelsum}{\mathbin{\!/\mkern-5mu/\!}}

\setcitestyle{super}

\begin{document}

\title{Enhancing SPDC brightness using elliptical pump shapes}
\author{Aitor Villar}
\affiliation{%
 Centre for Quantum Technologies, National University of Singapore, 3 Science Drive 2, S117543\\
}%
\author{Arian Stolk}%
\affiliation{%
 Centre for Quantum Technologies, National University of Singapore, 3 Science Drive 2, S117543\\
}%
\affiliation{Currently with QuTech, Delft University of Technology, PO Box 5046, 2600 GA Delft, The Netherlands}
\author{Alexander Lohrmann}%
\affiliation{%
 Centre for Quantum Technologies, National University of Singapore, 3 Science Drive 2, S117543\\
}%
\author{Alexander Ling}
\affiliation{%
 Centre for Quantum Technologies, National University of Singapore, 3 Science Drive 2, S117543\\
}%
 \affiliation{Physics Department, National University of Singapore, 2 Science Drive 3, S117542}

\date{\today}

\begin{abstract}
We report on the use of elliptical pump spatial modes to increase the observed brightness of spontaneous parametric downconversion in critically phase matched crystals. Simulations qualitatively predict this improvement which depends on the eccentricity and orientation of the pump ellipse. We experimentally confirm a factor of two improvement in brightness when compared to the traditional circular-symmetric pump spatial modes. These results support previous theoretical work that proposes the use of elliptical pump modes to enhance the performance of parametric processes in anisotropic materials.
\end{abstract}

\maketitle



Entangled photon pair sources play a crucial role in fundamental tests of nature \cite{bouwmeester1997experimental,yin2017satellite} and emerging quantum applications \cite{poppe2004practical,bell2013multicolor}. Spontaneous parametric downconversion\cite{burnham1970observation} (SPDC) is the most common method to generate these photon pairs, with a variety of designs\cite{kwiat1999ultrabright,fedrizzi2007wavelength,villar2018experimental} and materials used. In SPDC, a photon (pump, $p$) in a nonlinear crystal can downconvert into two photons (signal, $s$; idler, $i$) obeying energy and momentum conservation.

\begin{figure}[!t]
\centering
	\includegraphics{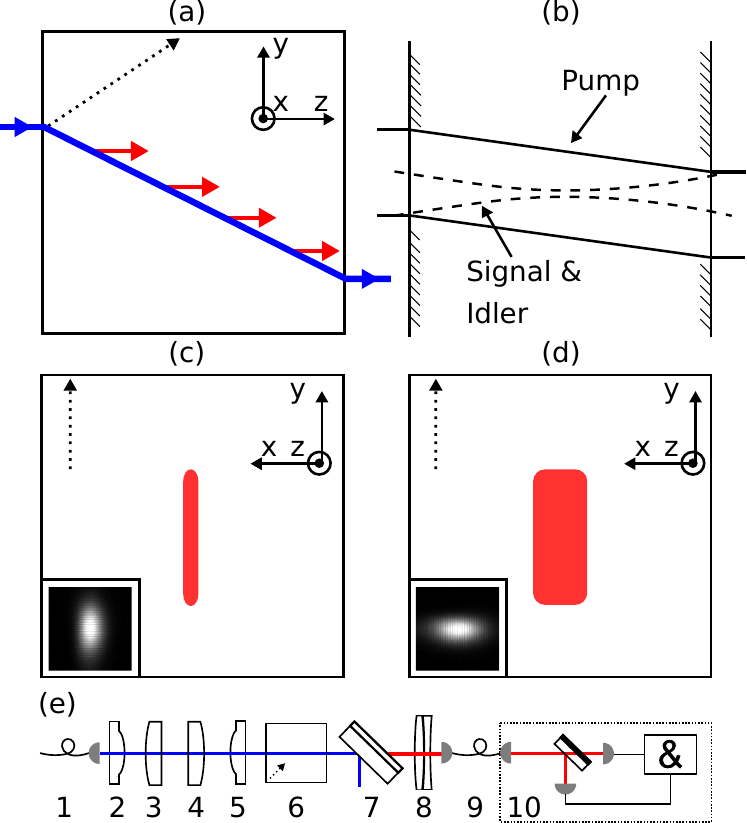}
\caption{(a) Side view of the pump spatial walk-off (blue line) within the crystal. The collinear SPDC emission is depicted by red arrows. The dotted line is the crystal's optical axis. 
(b) Sketch adapted from\cite{kuizenga1972optimum} representing the overlap between the pump beam (solid lines) and the downconversion beams (dotted lines), depicting the importance of controlling the pump focusing in order to maximize the overlap. (c) Collinear SPDC emission profile at the exit face of the crystal when the major axis of the pump ellipse (inset) is parallel to the walk-off direction. (d) Same as (c) but the ellipse major axis is rotated by 90$\degree$. (e) Experimental setup, 1: single-mode fiber pump output, 2: collimation lens, 3-4: cylindrical lenses, 5: focusing lens, 6: BBO crystal, 7-8: excess pump removal optics, 9: single-mode fiber, 10: photon separation via a dichroic mirror followed by coincidence detection setup.}
\label{fig:experiment}
\end{figure}

Early studies identified the influence of pump parameters on parametric brightness\cite{boyd1966theory}. The most comprehensive review on pump optimization was made by Boyd and Kleinman\cite{boyd1968parametric}, who suggested the use of circular symmetric spatial modes for pumping nonlinear processes. Though used most often, assumptions about circular (isotropic) symmetry might not optimize frequency conversion given that the vast majority of parametric processes take place in anisotropic materials. A specific case of interest, where the anisotropy is important, is when using critical phase matching in birefringent materials. When the pump is extraordinarily polarized it experiences spatial walk-off, as depicted in Fig.~\ref{fig:experiment}(a). The interplay of pump spatial modes with walk-off was first studied by Volosov\cite{volosov1970effect} and Kuizenga\cite{kuizenga1972optimum}. They showed that elliptical pump shapes could improve frequency conversion performance in second-harmonic generation (SHG) and other parametric processes. Their main observation was that the spatial walk-off limits the overlap of any parametric emission (see Fig.~\ref{fig:experiment}(b)). This restricts the benefits of tight pump focusing in the direction of walk-off. In contrast, this restriction is absent in the non-walk-off direction, which allows for tighter focusing. Hence, it is natural to consider different focusing conditions for different directions, leading to an elliptical (astigmatic) spatial mode.


\begin{figure*}[!t]
\centering
\includegraphics{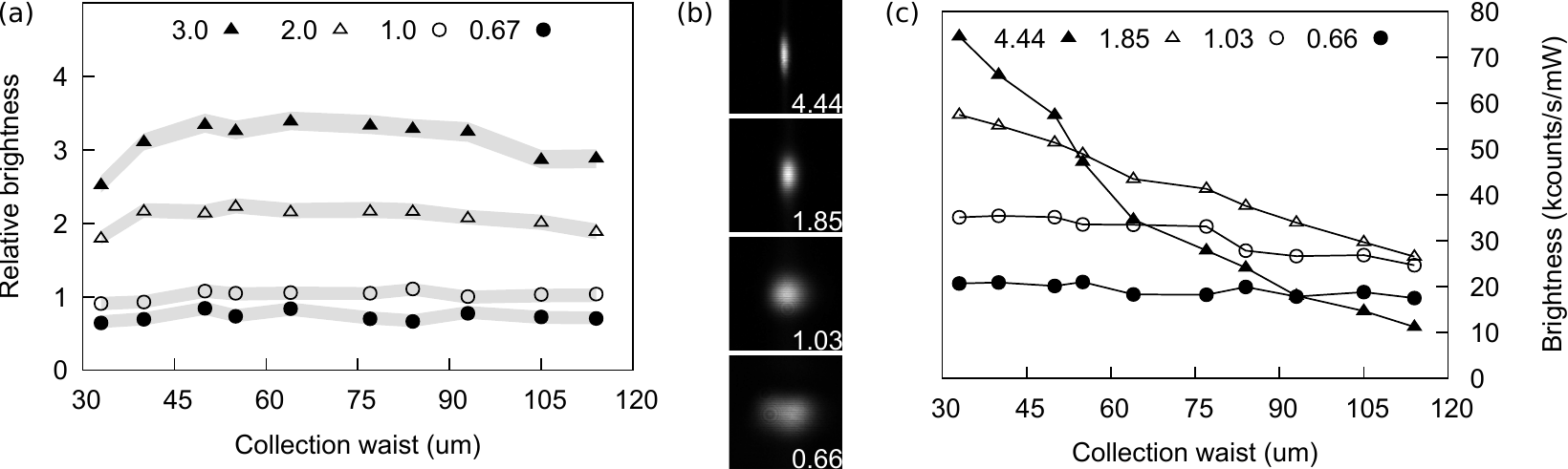}
	\caption{Results for the regime where the major axis of the pump ellipse is parallel to the walk-off direction. (a) Simulated in-fiber brightness at different collection distances for different pump aspect ratios. The shaded area indicates 1 s.d. confidence interval assuming Poisson statistics. (b) CCD camera images of the pump at the crystal position. (c) Experimentally observed brightness at different collection distances for different pump aspect ratios.}
\label{fig:results_1}
\end{figure*}
Experimental studies on SHG using elliptical pump beams verified the predicted gain, but the general observation was that the additional experimental complexity involving cylindrical focusing appeared to offset any enhancement\cite{steinbach1996cw,freegarde1997general}. This observation stems from the challenges of operating elliptical beams in a resonant cavity. However, single-pass parametric processes such as SPDC are exempt from these difficulties.

In this work, we study the effect of combining elliptical pump and circular-symmetric collection modes in type-I SPDC with negative birefringent materials. Our approach is outlined in Fig.~\ref{fig:experiment}. We distinguish between two pump orientations; parallel (Fig.~\ref{fig:experiment}(c)) and orthogonal (Fig.~\ref{fig:experiment}(d)) to the walk-off plane (defined by the propagation and walk-off direction).

To systematically investigate the effect of eccentricity the pump is launched from a single-mode fiber before undergoing cylindrical focusing. 
The eccentricity of the pump is defined via the aspect ratio $r=\frac{\omega_y}{\omega_x}$, where $\omega_y$ ($\omega_x$) is the vertical (horizontal) waist of the ellipse. In our discussion, the vertical direction is associated with the walk-off.
The change in pump aspect ratio was achieved by tuning only the waist of the horizontal axis.
For the ideal pump shape, Kuizenga suggests that the size of the pump in the two directions are determined separately by considering whether walk-off is present.

To optimize the pump shape in each direction, we should utilize the birefringence parameter $B$\cite{boyd1968parametric}:
\begin{equation}
B=\frac{\rho}{2}(lk_{0})^{1/2},
\label{eq:b_param}
\end{equation}
where $\rho$ is the walk-off angle, $l$ is the length of the crystal and $k_0$ is the pump wave vector. In the direction where there is no walk-off, $B=0$.

The experimental layout for investigating elliptical pump performance is shown in Fig.~\ref{fig:experiment}(e).
A \SI{5}{\milli\metre} $\beta$-Barium Borate (BBO) crystal (cut-angle $\theta = 28.76\degree$) is pumped with \SI{405}{\nano\metre} light to generate type-I, collinear, non-degenerate SPDC wavelengths (signal and idler wavelengths at \SI{776}{\nano\metre} and \SI{847}{\nano\metre}, respectively). 
Excess pump power was removed by a dichroic mirror and a long-pass filter. 
Finally, the SPDC photons were coupled into a single-mode fiber and sent to a detection setup.

A range of pump aspect ratios were realized with different cylindrical lenses. Once a target aspect ratio was achieved, a plano-convex lens focused down the pump ellipse. 

According to the literature \cite{boyd1968parametric}, the optimal pump waist in the walk-off direction (given our crystal properties) is \SI{22}{\micro\metre}, whereas the optimal pump waist in the direction with no walk-off is \SI{7.5}{\micro\meter}. This gives the optimal aspect ratio of $r=2.9$.

These optimal values, however, do not take into account experimental limitations, such as aberrations introduced by thick lenses. Empirically, we found that in our setup a pump waist of \SI{93}{\micro\metre} optimized the brightness and collection efficiency when the aspect ratio was 1.0, with a collection waist of \SI{45}{\micro\metre}.

Therefore, to investigate the effect of the pump aspect ratio we fixed the pump size in the walk-off direction at \SI{93}{\micro\metre}. Correspondingly, the pump size in the non-walk off direction ranged from \SI{27}{\micro\meter} to \SI{150}{\micro\meter}. Thus, the range of aspect ratios that were implemented ranged from 4.44 to 0.66. The experiment was then repeated for the case where the major axis of the pump beam was orthogonal to the walk-off direction.
In this case, the experimental aspect ratio investigated ranged from 2.94 to 0.65.

\begin{figure*}[!t]
	\centering
	\includegraphics{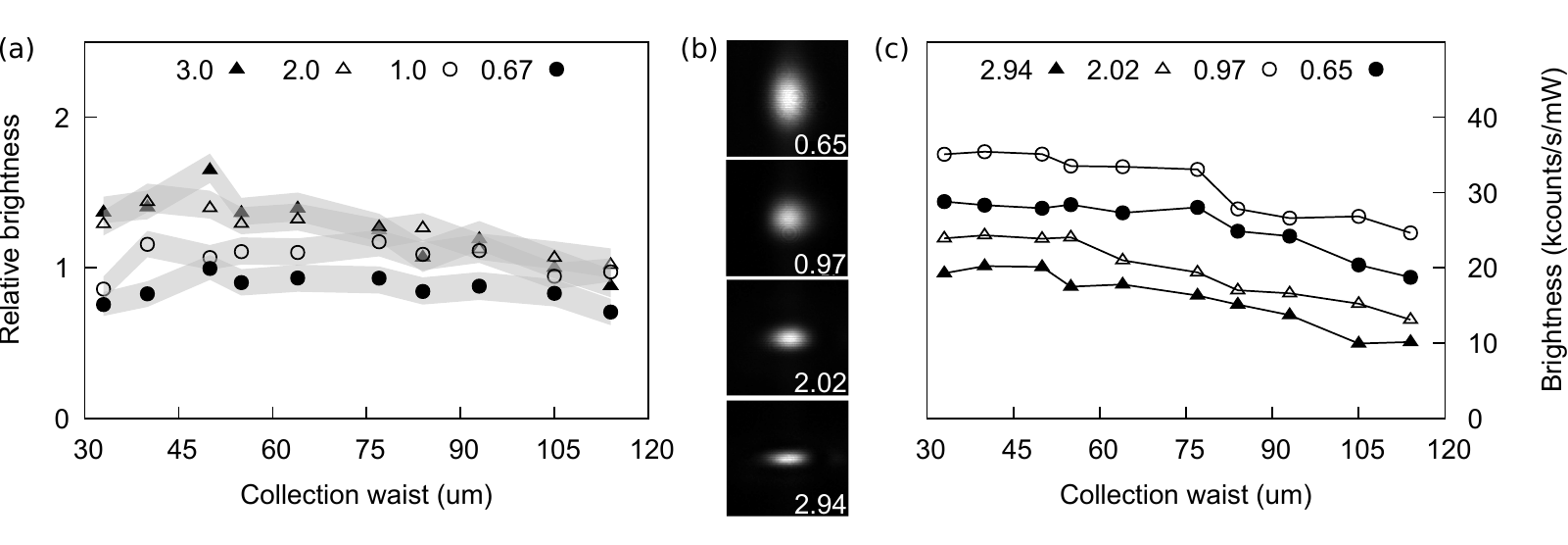}
	\caption{Results for the regime where the major axis of the pump ellipse is orthogonal to the walk-off direction. (a) Simulated in-fiber brightness at different collection beam waists for different pump aspect ratio. Here, the ellipse is rotated \SI{90}{\degree} with respect to the data presented in Fig.~\ref{fig:results_1}. The shaded area indicates 1 s.d. confidence interval assuming Poisson statistics. (b) CCD camera images of the pump at the crystal position. (c) Experimental in-fiber brightness at different collection distances for different pump aspect ratios.}
	\label{fig:results_2}
\end{figure*}

To provide better insight on the interplay of elliptical pump and circular collection modes a model was developed.
Within the model, the SPDC intensity is governed by the conventional phase matching conditions of the respective ($p$, $s$ and $i$) fields. 
The classical $p$ field is treated as a probability distribution of $\vec{k}_p$, that follows the angular and spatial intensity distributions of a Gaussian beam profile. 

For a given wavelength $\lambda_p$, starting position within the crystal $\{x,y,z\}$ and propagation direction $\vec{k}_{p}$, the weighting function $A$ can be obtained:
\begin{equation}
A= e^{-\frac{( \omega_{x} \Delta k_{x} )^2+( \omega_{y} \Delta k_{y})^2}{2}} \cdot\sinc\bigg(\frac{l\Delta k_{z}}{2}\bigg)^2 .
\label{eq:sinc}
\end{equation}
Here $\Delta \vec{k} = \Delta k_x\hat{x} + \Delta k_y \hat{y} + \Delta k_z \hat{z}$ is the phase mismatch in Cartesian coordinates, $\omega_{x,y}$ the beam waist in horizontal (\textit{x}) and vertical (\textit{y}) directions, and $l$ the crystal length. This weighting function determines the probability of generating a signal/idler pair with wavelengths $\lambda_{s}$ and $\lambda_{i}$ in the propagation directions $\hat{k}_{s}$ and $\hat{k}_{i}$. The individual treatment for each $\vec{k}_p$ accounts for effects of the wavefront curvature.

Using conventional ray-tracing techniques the $\vec{k}_{s},\vec{k}_{i}$ pair was propagated through the downstream optics towards a single-mode fiber (numerical aperture of $0.1$ and core diameter \SI{5}{\micro\metre}). The overlap of the SPDC rays with the collection mode of the fiber determines the in-fiber photon pair brightness. The model is available online at: \url{https://github.com/arianstolk/SPDC_test/}.

The simulated and experimental results are shown in Fig.~\ref{fig:results_1} and Fig.~\ref{fig:results_2}. 
When the major axis of the pump ellipse is parallel to the walk-off direction, the experimental data agrees qualitatively with the model, which predicts increased brightness for larger pump aspect ratio. The best observed improvement over the circular symmetric mode is a factor of 2.
When the major axis of the pump ellipse is orthogonal to the walk-off direction, there is a dramatic difference as the brightness shows little response when the aspect ratio changes.

In the case where the major axis is aligned with the walk-off direction, increasing the aspect ratio monotonically increased the photon pair brightness that can be observed. Due to experimental difficulties, the optimal aspect ratio of 2.9 was not achieved. However, it is instructive to consider the results of the two aspect ratios that bracket this value. These results showed a strong dependence of brightness on the orientation of the pump major axis, validating Kuizenga's proposal.

In Fig.~\ref{fig:results_1}(c), there is a strong linear relationship between the brightness and collection waist, which is not present in the model.This is attributed partly to the lack of precision within the model which uses the thin lens approximation for modeling the single-mode fiber photon collection and ignores the lens thickness. Furthermore, the observed $M^2$ value of the pump beam in the walk-off direction deviates from the ideal case for large aspect ratio (see Tab.~\ref{tab:params} in Appendix) due to lens aberrations when using thick cylindrical lenses. 

Improvements in the model (e.g. improved approximations for ray tracing of the curved optics) will help to match the experimental results and provide better predictions about the overall source performance. 
Additionally, an extension of this work would be the use of elliptical collection modes that are tailored to the pump profile and crystal anisotropy.

When considering asymmetric effects of the phase matching conditions on the SPDC profile, the use of a correctly oriented elliptical pump mode showed improved in-fiber brightness over the traditional circular symmetric modes. This deviates from experimental results in SHG in optical parametric oscillators, where the marginal improvement in performance did not warrant the expense of increased experimental complexity.  Elliptical pump modes have already improved the brightness of entangled photon pairs sources where the crystal geometry allows an elliptical pump mode to be implemented\cite{villar2018experimental,lohrmann2018high}.

In many modern SPDC sources the pump is typically obtained from a laser diode which inherently exhibits an elliptical profile.  
Taking into account crystal anisotropy and beam walk-off it might be unnecessary to correct the elliptical pump profile, simplifying the construction of photon pair sources while improving performance.

\section*{Acknowledgments}
This research is supported by the National Research Foundation, Prime Minister's Office, Singapore under its Competitive Research Programme (CRP Award No. NRF-CRP12-2013-02). This program was also supported by the Ministry of Education, Singapore.
\bibliography{bibliography}
\clearpage
\onecolumngrid
\begin{appendices}
\section{The Boyd-Kleinman optimal circular symmetric pump beam calculation}
In order to calculate the optimal circular symmetric pump beam suggested by Boyd-Kleinman, first the $B$ parameter of the nonlinear crystal needs to be calculated. In our case, the properties of the BBO crystal: $\rho=\SI{3.83}{\degree}$, $l=\SI{5}{\milli\metre}$ and $\lambda_0 = \SI{405}{\nano\metre}$. From Eq.~\ref{eq:b_param} in the main text, $B=9.3$.

Knowing the $B$ parameter, the optimal $\xi$ value can be inferred via the optimization curves shown in Fig.~\ref{fig:bk-curves}:

\begin{figure}[h]
	\centering
	\includegraphics{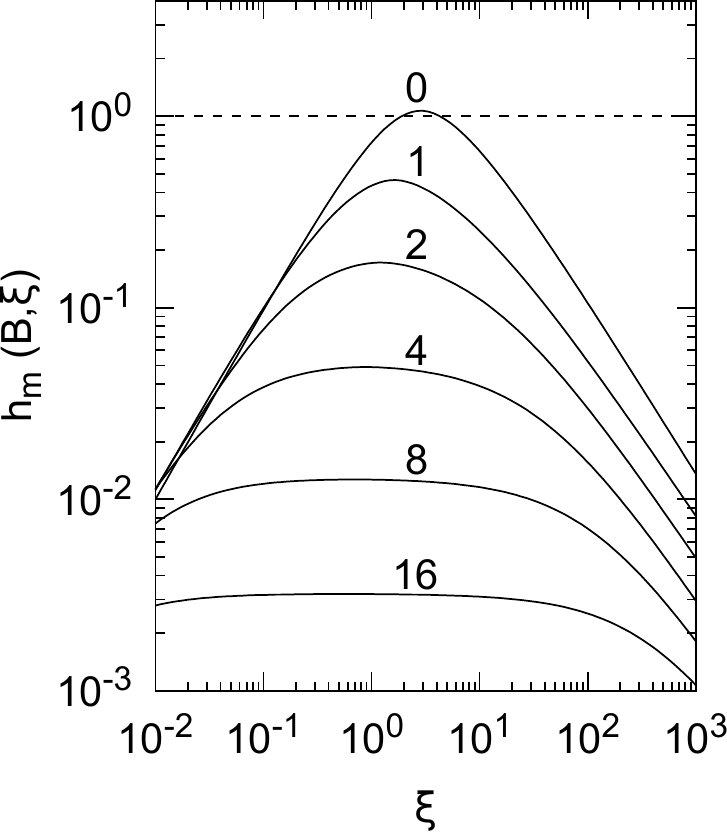}
	\caption{Optimization curves for circular symmetric beams reported in \cite{boyd1968parametric}. For different walk-off parameters (the $B$ number is shown above each curve) $\xi$ (x-axis) and the figure of merit $h_m$ (y-axis) is shown. Resorting to this curves and knowing the walk-off angle and the thickness of the crystal, the optimal circular symmetric beam for a particular wavelength $\lambda_0$ can be determined.}
	\label{fig:bk-curves}
\end{figure}

For $B=9.3$, $\xi$ is calculated to be 0.6579. From here, the ideal beam waist can be calculated, since $\xi=\frac{l}{b}=\frac{l\cdot\lambda}{\omega^2 \cdot 2\pi}$. In our case, $\omega=\SI{22.13}{\micro\metre}$.

\section{Pump beam parameters}
Pump beam parameters used in both simulation (\textit{Sim}) and experiment (\textit{Exp}) are depicted in Tab.~\ref{tab:params}. Due to experimental factors (e.g., lens aberrations) when shaping the pump mode, the experimental Rayleigh length values are lower than the ideal values used in the simulation ($M^{2}_{\perp},M^{2}_{\parallelsum} = 1$). This explains the difference in decaying rates between simulation and experiment.
\begin{table*}[htb]
	\centering
	\caption{The Rayleigh length is defined as $Z_R = \frac{\pi \omega_0^{2}}{M^{2}\lambda_p}$, where $\omega_0$ is the beam waist at the focus position, $M^{2}$ is the beam quality factor and $\lambda_p$ is the pump wavelength. The $\parallelsum$ ($\perp$) direction is defined as the direction parallel (orthogonal) to the phase matching plane.}

\begin{tabular}{|c| c | c | c | c | c | c | c | c |}
      \hline
      \multirow{2}{*}{Pump Param.}& \multicolumn{2}{c|}{Aspect ratio} & \multicolumn{2}{c|}{$ Z_{R\parallelsum}$ [mm] }&$M^{2}_{\parallelsum}$ &\multicolumn{2}{c|}{$Z_{R\perp}$[mm]} &$M^{2}_{\perp}$ \\\cline{2-9}       							
      &Sim&Exp&Sim&Exp&Exp&Sim&Exp&Exp\\\hline\hline
      \multirow{4}{*}{Parallel} & 0.67 & 0.66 & 174.5 & 145.6 &1.1& 77.6 & 69.1&1.1\\\cline{2-9} 
      
       								& 1.0 & 1.03 & 77.6& 65.2 &1.1 & 77.6 &70.9 &1.1 \\\cline{2-9} 
      
      								& 2.0 &1.85 & 19.4& 20.9 &1.0 & 77.6 &49.1&1.5 \\\cline{2-9}  
      
       								& 3.0 & 4.44 & 8.4& 5.0 &1.0& 77.6& 30.8 &1.7\\\cline{2-9}

      \hline\hline
      \multirow{4}{*}{Orthogonal} & 0.67 & 0.65 & 77.6 & 76.0 &1.0& 174.5 & 155.5&1.1 \\\cline{2-9}  
      
    								& 1.00& 0.97 & 77.6 & 65.2&1.1 & 77.6& 70.9&1.1 \\\cline{2-9}  
      
    								& 2.0 &2.02 & 77.6& 66.2&1.1 & 19.4 &17.3&1.2 \\\cline{2-9} 
      
    								& 3.0 &2.94 & 77.6 &70.9&1.0 & 8.4 &8.2&1.0 \\\cline{2-9}

  \hline
     
\end{tabular}     
\label{tab:params}
\end{table*}

\end{appendices}

\end{document}